\documentclass[pra, aps, 10pt, superscriptaddress, twocolumn]{revtex4-1}

\usepackage{hyperref}

\usepackage[utf8]{inputenc}
\usepackage{lmodern}
\usepackage[version=3]{mhchem}
\usepackage{amsmath,amssymb,amstext,amsfonts}
\usepackage{natbib}
\usepackage{psfrag,graphicx}
\usepackage[]{units}
\usepackage{upgreek}
\usepackage{textcomp} 
\usepackage{color}
\usepackage{xcolor}
\usepackage{float}
\usepackage{subfigure}
\usepackage{array} 
\usepackage{epstopdf}
\usepackage[ngerman, num]{isodate}

\begin{document}
\title{Coercivity enhancement of selective laser sintered NdFeB magnets by grain boundary infiltration}

\author{Christian Huber}
\thanks{Correspondence to: \href{mailto:huber-c@univie.ac.at}{huber-c@univie.ac.at}}
\affiliation{Physics of Functional Materials, University of Vienna, 1090 Vienna, Austria}
\affiliation{Christian Doppler Laboratory for Advanced Magnetic Sensing and Materials, 1090 Vienna, Austria}

\author{Hossein Sepehri-Amin}
\affiliation{National Institute for Materials Science, Tsukuba 305-0047, Japan}

\author{Michael Goertler}
\affiliation{Institute for Surface Technologies and Photonics, Joanneum Research Forschungsgesellschaft GmbH, 8712 Niklasdorf, Austria}

\author{Martin Groenefeld}
\affiliation{Magnetfabrik Bonn GmbH, 53119 Bonn, Germany}

\author{Iulian Teliban}
\affiliation{Magnetfabrik Bonn GmbH, 53119 Bonn, Germany}

\author{Kazuhiro Hono}
\affiliation{National Institute for Materials Science, Tsukuba 305-0047, Japan}

\author{Dieter Suess}
\affiliation{Physics of Functional Materials, University of Vienna, 1090 Vienna, Austria}
\affiliation{Christian Doppler Laboratory for Advanced Magnetic Sensing and Materials, 1090 Vienna, Austria}

\date{\today}


\begin{abstract}
Laser powder bed fusion is a well-established additive manufacturing method that can be used for the production of net-shaped Nd-Fe-B sintered magnets. However, low coercivity has been one of the drawbacks in the laser powder bed fusion processed Nd-Fe-B magnets. In this work, we have demonstrated that the grain boundary diffusion process using low-melting Nd-Cu, Nd-Al-Ni-Cu, and Nd-Tb-Cu alloys  to the selective laser sintered NdFeB magnets can results in a substantial enhancement of coercivity from $0.65$~T to $1.5$~T.  Detailed microstructure investigations clarified formation of Nd-rich grain boundary phase, introducing Tb-rich shell at the surface of Nd$_{2}$Fe$_{14}$B grains, and maintaining the grain size in nano-scale are responsible for the large coercivity enhancement.
\end{abstract}

\maketitle

\section*{Introduction}
Additive manufacturing (AM) or colloquially called 3D printing of permanent magnetic materials offer new possibilities in advanced applications through the design freedom of the magnetic parts. The object is fabricated in a layer-by-layer manner of formless of form-neutral construction material by means of thermal or chemical processes. Magnets manufactured by AM methods have several advantages compared to traditional manufacturing methods like sintering or injection-molding of polymer-bonded magnets. Advantages include net shape capability, reduced waste, and mechanical flexibility, etc. Fused Deposition Modelling (FDM) is a reasonable and rapid AM method to process polymer-bonded NdFeB magnetic material \cite{pub_16_1_apl, pub_17_2, pub_17_1, ortner2017application, baam, von20183d}. Disadvantages of FDM hard magnets are the limited operational temperature of the polymer-matrix material and the reduced $(BH)_\text{max}$ compared to sintered magnets due to the isotropic nature of FDM hard magnets. Binder jetting of NdFeB powder was successfully shown in \cite{binder_jetting}. Disadvantage of this method is the low density of only $50$~\% compared to the full-dense object. Nevertheless, this porous 3D printed structures are well suited for an alloy infiltration process to increase the density and coercivity of the printed magnets.  Low-melting point eutectic alloys (i.e. NdCuCo, PrCuCo) can be used to infiltrate a structure and double the coercivity \cite{li2017novel}.

Laser powder bed fusion (LPBF) is well established method to produce sintered and fully dense metallic objects \cite{slm}. It completely melts the metal powder by the aid of a high-power laser source. To increase the quality of the printed structures and its mechanical properties, powders with a spherical morphology are required \cite{attar2015effect}. Soft magnetic materials can be in-situ synthesized with this method \cite{huber2018additive, mikler2017laser, zhang2013magnetic}. LPBF of hard magnetic AlNiCo powder with a spherical morphology is presented in \cite{white2017net}. Most recent publications in the field of LPBF of hard magnets use a commercial NdFeB powder (MQP-S-11-9 supplied by Magnequench Corporation) \cite{lasersinter_mag, urban2017influences}. This grade has a spherical morphology and its main field of application is the manufacturing of bonded magnets, particularly by injection molding, extrusion and calendering. It is based on a patented NdPrFeCoTiZrB alloy \cite{kanekiyo2004nanocomposite}. The nano-sized NdFeB grains have an uniaxial magnetocrystalline anisotropy, and the orientation of the grains is random leading to isotropic magnetic properties of the bulk magnet. It is produced by employing an atomization process followed by heat treatment

As mentioned before, the process parameters of the LPBF process are usually set to completely melt the powder material locally by means of a laser source, forming a solid material layer after solidification. This rapid liquefaction and the subsequent cooling down of the melt, influences the size of the grains and the composition of the grain boundaries \cite{fischer1996grain, engelmann1997microstructure, kronmuller1988analysis}. This modification of the microstructure affect the coercivity of the additive manufactured magnets. High performance motors and generators have an operating temperature of around 200~$^\circ$C, and due to the thermal coercivity degeneration of NdFeB, the coercivity at room temperature should be maximized \cite{hono2012strategy}. Therefore, we refrain from completely melting each powder layer but sinter the particles, to retain their original microstructure. First, doping with Dy increase the coercivity of NdFeB up to $3$~T, but the magnetization decrease substantially \cite{boltich1985magnetic, hirosawa1986magnetization}. Alternatively, the coercivity of NdFeB magnets can be substantially enhanced by diffusion of Nd-Tb-Cu and Nd-Al-Ni-Cu eutectic alloy through grain boundaries (GB) \cite{liu2016coercivity, li2018coercivity}

In this publication, we present a grain boundary (GB) infiltration method to increase the coercivity of LPBF manufactured samples. Several infiltration alloys are investigated, and detailed microstructural analyses investigate the GB diffusion processes.  

\section*{Experiments}
For sample fabrication a commercial Farsoon~FS121M LPBF-machine was used. It is equipped with a continuous wave $200$~W Yb-fibre laser with a wavelength of $1.07$~\textmu m and a spot size of $0.1$~mm. The spherical MQP-S-11-9-20001 powder produced by Magnequench Corporpation was used in this study \cite{mqp}. In \cite{lasersinter_mag}, the characteristics of the powder were investigated: the chemical composition states Nd$_{7.5}$Pr$_{0.7}$Fe$_{75.4}$Co$_{2.5}$B$_{8.8}$Zr$_{2.6}$Ti$_{2.5}$ (at.\%), the powder size distribution shows a d$_{50}$ of $38$~\textmu m and the tap density exhibits $61.4$~\% of the materials full density. Since the MQP-S powder exhibits a lower Nd-content than the Nd$_{2}$Fe$_{14}$B hard magnetic phase, its magnetic properties are not superior.

Considering its powder size distribution and morphology, the MQP-S powder meets the requirements of the recoating process during LPBF processing \cite{debroy2018additive}. The morphology of the MQP-S powder can be seen in the scanning electron microscopy with backscattered electron contrast (BSE-SEM) image shown in Fig.~\ref{fig:se_sem_microstructure}(a). The vast majority of particles is spherical with no or few small satellites and a minor fraction of particles is rod-like. Some powder particles are broken indicating brittle material behavior. Studies of Jaćimović et al. and Kolb et al. document the production of magnetic samples by fully melting the MQP-S powder to cubic specimens in an LPBF machine \cite{lasersinter_mag, kolb2016laser}.

In this study, we plan to additively manufacture specimens from the MQP-S powder and infiltrate them with low-melting point eutectic alloys in order to investigate the effect of the infiltration process on the magnet’s coercivity. Since a high surface to volume ratio enables an efficient diffusion of the alloys through the grain boundaries, we produced specimens with a micro-porous inner structure.  This is achieved by only sintering and not fully melting the MQP-S powder during the LPBF process. Therefore, the process parameters of the LPBF machine were adjusted accordingly.

The printing was performed under argon atmosphere with oxygen content below $0.1$~\%, a layer thickness of $20$~\textmu m and the powder recoating was done with a carbon fiber brush. All specimens were printed without support structured directly onto a steel substrate plate to ensure proper heat dissipation.

To find a proper parameter regime, melt track studies were performed. For the production of single melt lines the parameters laser power $P$ and scan speed $v$ were varied according to a matrix: $P$ was varied between $20$~W and $100$~W and $v$ between $50$~mm/s and $2000$~mm/s.  To minimize the effect of chemical mixing between the MQP-S powder and the steel substrate during laser exposure, five lines were printed stacked upon each other with a layer thickness of $20$~\textmu m. For interpretation purposes, we introduce the artificial magnitude line energy $E_\text{line}$, which reads as $P$ over $v$. As a result, completely molten tracks were produced for $P \geq 40$~W and $E_\text{line} \geq 0.1$~J/mm. Since full melting of the powder is improper for sintering, we focused on line energies between $0.03$~J/mm and $0.07$~J/mm at $40$~W for the production of volumetric specimens.

For the production of volumetric specimens, parallel melt tracks are produced next to each other with a defined offset, called hatch spacing $h$. Together with the above mentioned parameter combinations of $E_\text{line}$ and $P$, $h$ was varied between $0.1$~mm and $0.14$~mm to find proper parameter sets for the production of intactly sintered cubes with dimensions of $5\times5\times5$~mm$^3$. The produced specimens were inspected optically concerning their structural integrity. Intact cubes could be produced for volume energy inputs $E_\text{vol}$ between $0.14$ and $0.21$~J/mm$^3$. The volume energy input reads as $E_\text{line}$ over the layer thickness and hatch distance.

The average density of the cubes was determined to $65 \pm 2$~\% by dry weighing each cube, which is in good accordance with the MPQ-S powder’s tap density of $61.4$~\% as measured by \cite{lasersinter_mag}. This shows that the MPQ-S powder was sintered without complete melting of the material. 

Low melting point melt-spun ribbons with compositions of Nd$_{70}$Cu$_{30}$, Nd$_{80}$Cu$_{20}$, Nd$_{60}$Al$_{10}$Ni$_{10}$Cu$_{20}$, and Nd$_{50}$Tb$_{20}$Cu$_{20}$ were prepared by single role melt-spinner. The ribbons were crashed into the flake and the surface of LPBF processed Nd-Fe-B magnets were covered by the flakes with the mass ratio of $10$ to $1$ followed by annealing at a temperature of $650$~$^\circ$C for $3$~h under vacuum. The magnetic properties of the LPBF processed Nd-Fe-B magnets and diffusion processed magnets were measured using a superconducting quantum interface device vibrating sample magnetometer (SQUID-VSM). Microstructure of the samples were studied using Carl Zeiss CrossBeam 1540EsB. Transmission electron microscopy (TEM) was performed using a Titan G2 80-200 TEM with a probe aberration corrector. Energy dispersive X-ray spectroscopy (EDS) was collected using a Super-X EDX detector.

\section*{Results}
\begin{figure}[ht]
	\centering
	\includegraphics[width=1\linewidth]{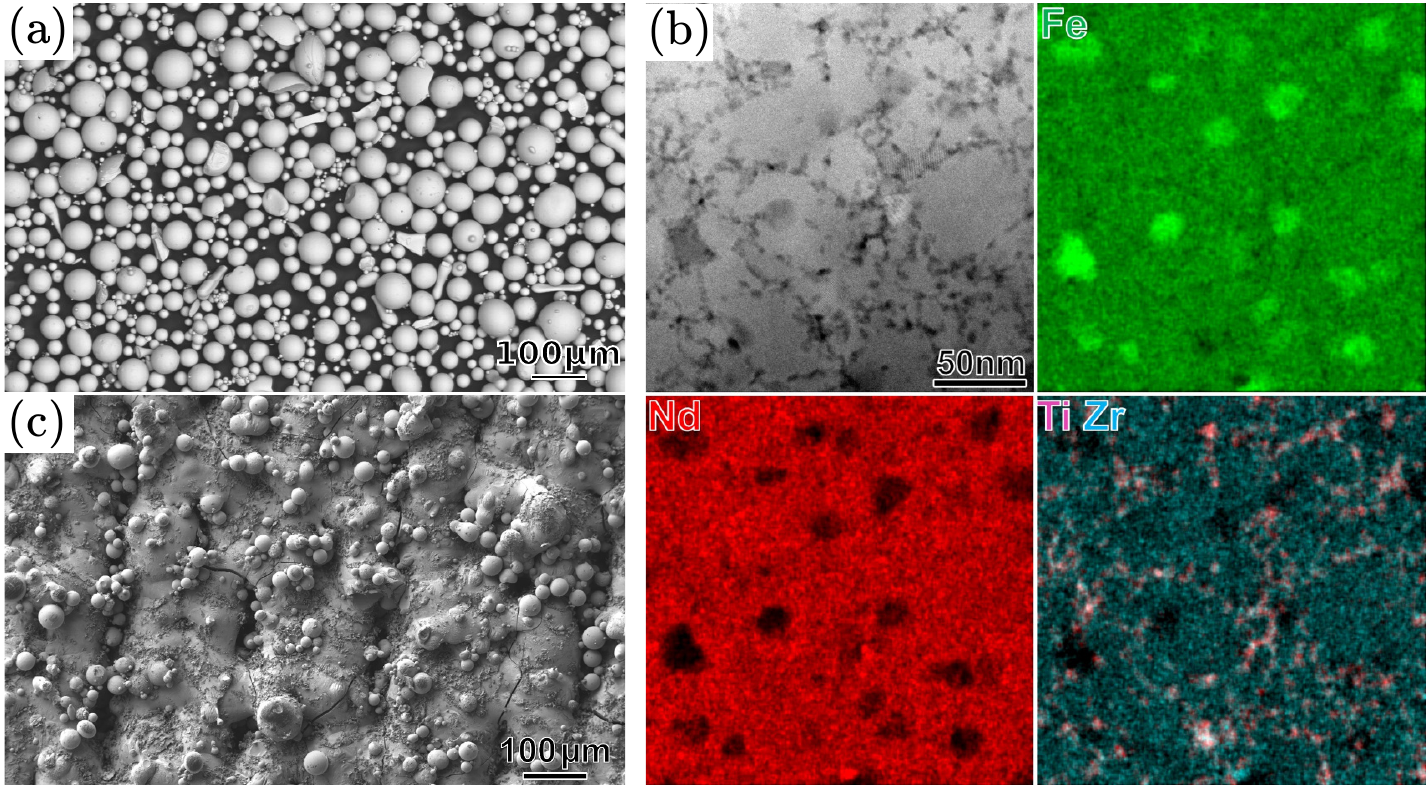}
	\caption{Microstructure of the LPBF processed initial sample. (a) SE-SEM image of the initial MQP-S powder. The major fraction of particles exhibits spherical morphology with no or small satellites and a minor fraction exhibits rod like morphology. (b) STEM-EDS maps of Fe, Nd, Ti, and Zr showing the microstructure of the gas atomized powders. (c) SEM image of the surface. Spherical particles with a diameter of $10-80$~\textmu m are visible. Some regions are porous and several cracks are observed.}
	\label{fig:se_sem_microstructure}
\end{figure}

Fig.~\ref{fig:se_sem_microstructure}(a) shows secondary electron (SE) SEM image of the gas atomized powders used for LPBF process. The spherical shaped powder particles have a size of smaller than $100$~\textmu m. Fig.~\ref{fig:se_sem_microstructure}(b) shows high angle annular dark field (HAADF)-STEM image and STEM-EDS maps of Fe, Nd, Ti, and Zr showing microstructure of the gas atomized powders. The powders contain Nd$_{2}$Fe$_{14}$B grains with a size of smaller than $100$~nm along with Ti and Zr-rich phases. No distinct segregation of Nd can be seen at the grain boundaries of the powders and nano-sized $\alpha$-Fe can be seen in the microstructure of the starting powders.  This is due to the composition of the powders that contains total RE of $8.2$~at.\%, smaller than the stoichiometric composition of Nd$_{2}$Fe$_{14}$B phase.  SEM-SE image obtained from surface of LPBF sample is shown in Fig.~\ref{fig:se_sem_microstructure}(c) indicating densified sample while several cracks can still be seen in the microstructure.

\begin{figure}[ht]
	\centering
	\includegraphics[width=1\linewidth]{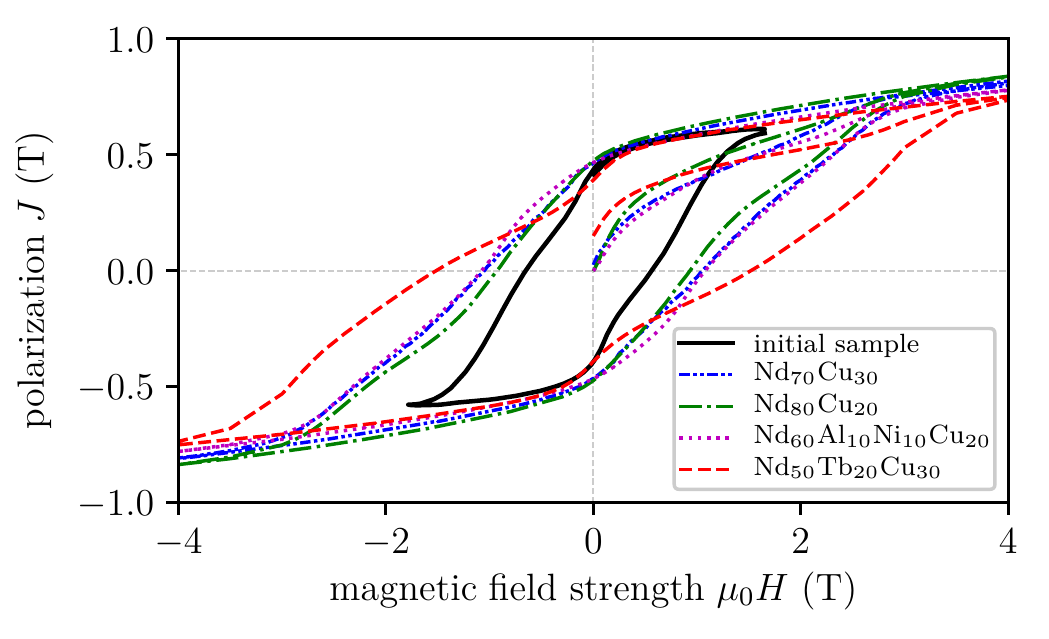}
	\caption{Hysteresis loops for the initial LPBF sample and for the diffusion-processed samples.}
	\label{fig:hysteresis}
\end{figure}

Fig.~\ref{fig:hysteresis} shows the magnetization curves of the initial LPBF processed magnet and diffusion processed samples using Nd$_{70}$Cu$_{30}$, Nd$_{80}$Cu$_{20}$, Nd$_{60}$Al$_{10}$Ni$_{10}$Cu$_{20}$, and Nd$_{50}$Tb$_{20}$Cu$_{20}$ diffusion sources. The coercivity of the initial LPBF processed sample is only $0.65$~T. This low coercivity is due to the absence of the Nd-rich grain boundary phase separating nano-sized Nd$_{2}$Fe$_{14}$B grains as well as the existence of soft $\alpha$-Fe phase in the microstructure as  shown in Fig.~\ref{fig:se_sem_microstructure}(b). Diffusion of Nd-Cu alloys and Nd-Al-Ni-Cu alloys increases the coercivity from $0.65$~T to $1.0$~T. However, use of Nd-Tb-Cu as the diffusion source results in achieving a larger coercivity of $1.5$~T in the samples. The samples after diffusion process showed a full density of above $7.5$~g/cm$^3$. Tab.~\ref{tab:mag_prop} shows the magnetic properties of all investigated samples.

\begin{table}[ht]
\caption{Magnetic properties of the initial and the diffusion-processed MQP-S-11-9 from Magnequench Corporation.}
\label{tab:mag_prop}
\begin{tabular}{l|ll}
sample                                  & $B_r$ (mT) & $\mu_0 H_{cj}$ (T) \\ \hline
powder (data sheet)                     & 746        & 0.880              \\
initial                                 & 436        & 0.653              \\
Nd$_{70}$Cu$_{30}$                      & 464        & 1.053              \\
Nd$_{80}$Cu$_{20}$                      & 475        & 0.973              \\
Nd$_{60}$Al$_{10}$Ni$_{10}$Cu$_{20}$    & 466        & 1.078              \\
Nd$_{50}$Tb$_{20}$Cu$_{30}$             & 390        & 1.518             
\end{tabular}
\end{table}

\begin{figure}[ht]
	\centering
	\includegraphics[width=1\linewidth]{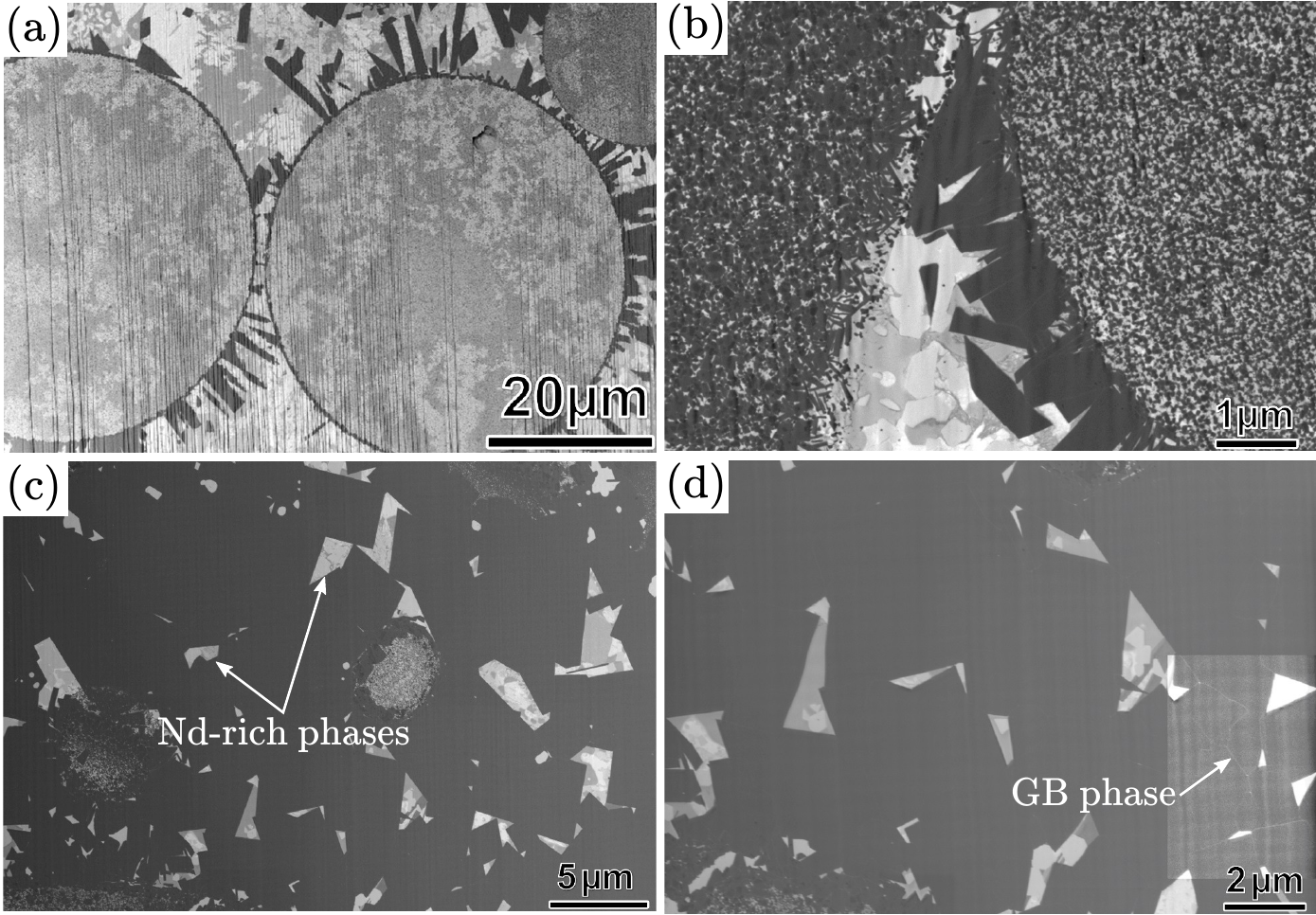}
	\caption{BSE-SEM image of the LPBF and with Nd$_{80}$Cu$_{20}$ diffusion-processed sample. (a) Different grain size inside the particles and at the interfaces. (b) Magnetic isolation of nano-sized Nd$_{2}$Fe$_{14}$B grains inside of the initial particles. (c) Dark gray regions correspond to grains of the NdFeB phase. Bright regions are so-called ``Nd-rich'' phases. (d) Grain boundaries are observed with inhomogeneous contrast.}
	\label{fig:nd_cu_diffusion_microstructure}
\end{figure}

Fig.~\ref{fig:nd_cu_diffusion_microstructure} shows backscattered electron (BSE) SEM images obtained from the  Nd$_{80}$Cu$_{20}$ diffusion processed sample. In this image, the brightly imaged contrast corresponds to the Nd-rich phase and the grayly imaged phase corresponds to the Nd$_{2}$Fe$_{14}$B matrix. Fig.~\ref{fig:nd_cu_diffusion_microstructure}(a) indicates the Nd-Cu flakes melt and infiltrate into the LPBF processed sample during the heat treatment at $650$~$^\circ$C filling the voids between the powder particles as well as diffusing inside the particles through the grain boundaries. SEM BSE image with a higher magnification is shown in Fig.~\ref{fig:nd_cu_diffusion_microstructure}(b) indicating formation of Nd-rich intergranular phases isolating nano-sized Nd$_{2}$Fe$_{14}$B grains. However, the interface of the powders shows a different microstructure, in particular different grain size, that is shown more clearly in Fig.~\ref{fig:nd_cu_diffusion_microstructure}(c) and (d). Although the size of Nd$_{2}$Fe$_{14}$B grains at the interface of the initial powders in LPBF processed magnet is as large as $2-5$~\textmu m, Nd-rich phase infiltrates through the grain boundaries forming Nd-rich grain boundary phases. Observed inhomogeneity in the Nd-Cu diffusion processed LPBF processed magnet is due to the effect of laser during the process. Laser heats and melts the surface of the gas atomized powders resulting in sintering of the surface of the powders. Hence, during slow solidification of the interface area, large sized Nd$_{2}$Fe$_{14}$B grains form at the interface of initial powders explaining observed heterogeneous grain size from the interface toward the center of the original powders in the LPBF processed magnet.

\begin{figure}[ht]
	\centering
	\includegraphics[width=1\linewidth]{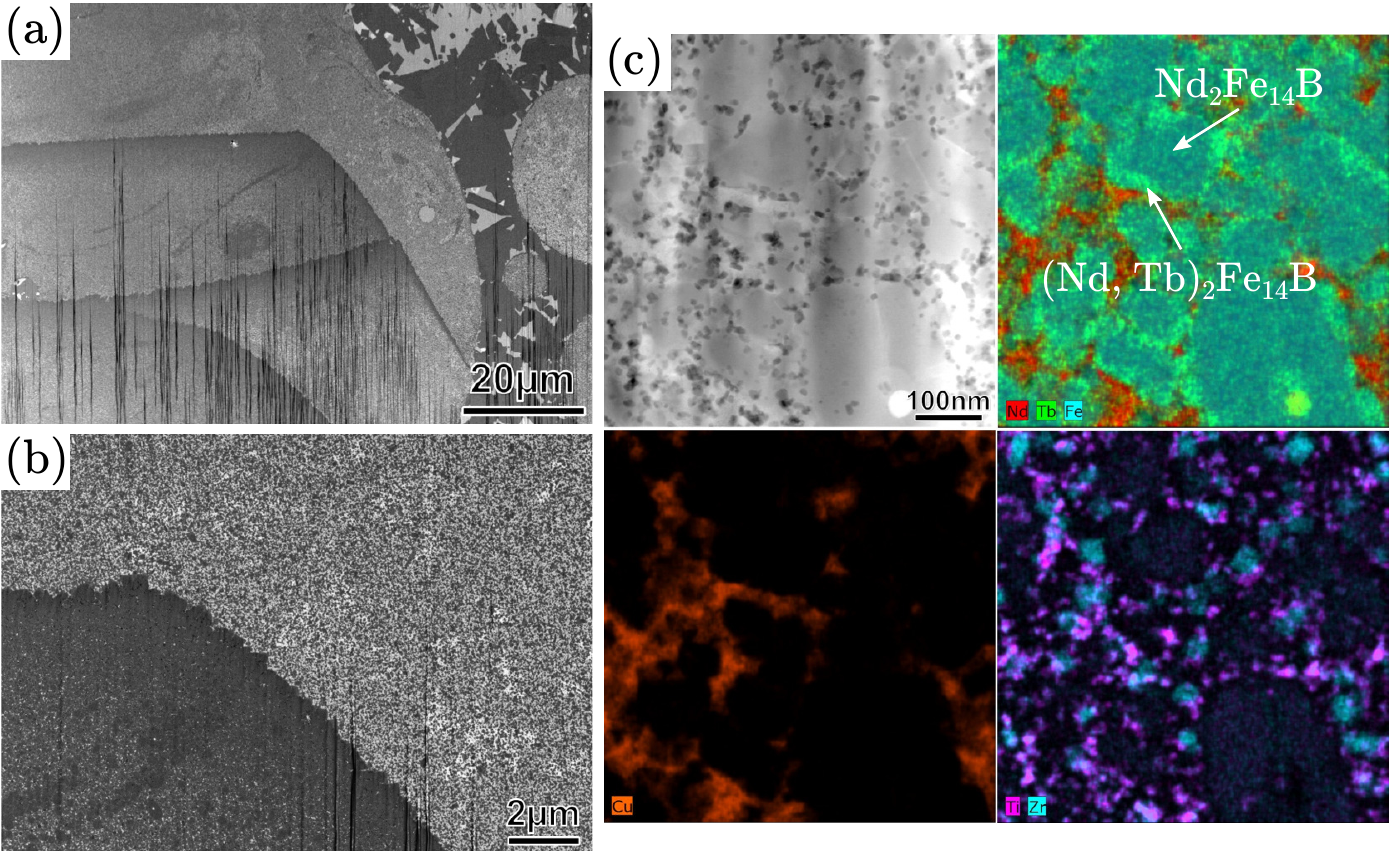}
	\caption{Microstructure of the LPBF and with Nd$_{50}$Tb$_{20}$Cu$_{30}$ diffusion-processed sample. (a) BSE-SEM image of the microstructure inside the grains and the interfaces. (b) BSE-SEM image indicates a diffusion of Nd-rich phase into the GB boundaries. (c) TEM-EDS elemental mapping images of Nd,Tb,Fe; Cu; Ti,Zr. A formation of a Nd-rich GB phase and a Tb-rich shell are visible. The initial particle grains are nano-sized and maintaining the 2:14:1 ratio.}
	\label{fig:nd_tb_cu_diffusion_microstructure}
\end{figure}

Fig.~\ref{fig:nd_tb_cu_diffusion_microstructure}(a) and (b) show low and high magnification SEM BSE images of Nd-Tb-Cu diffusion processed samples. Microstructure of this sample also shows a heterogeneous Nd$_{2}$Fe$_{14}$B grain size distribution, a large sized grains at the interface of initial powder particles and nano-sized grains inside of the particles that are isolated with Nd-rich intergranular phase, similar to that of observed in the Nd-Cu diffusion processed sample. High magnification HAADF-STEM image and STEM-EDS maps of Nd, Tb, Fe, Cu, Ti, and Zr are shown in Fig.~\ref{fig:nd_tb_cu_diffusion_microstructure}. Note that the Nd-rich phase is formed at the grain boundaries of Nd-Tb-Cu diffused sample, covering nano-sized Nd$_{2}$Fe$_{14}$B grains. A Tb-rich shell can be seen covering the surface of Nd$_{2}$Fe$_{14}$B grains.  No $\alpha$-Fe is observed in the microstructure that should be due to the reaction of Nd-Tb-Cu liquid phase with Fe phase during the diffusion process.

\begin{figure}[ht]
	\centering
	\includegraphics[width=1\linewidth]{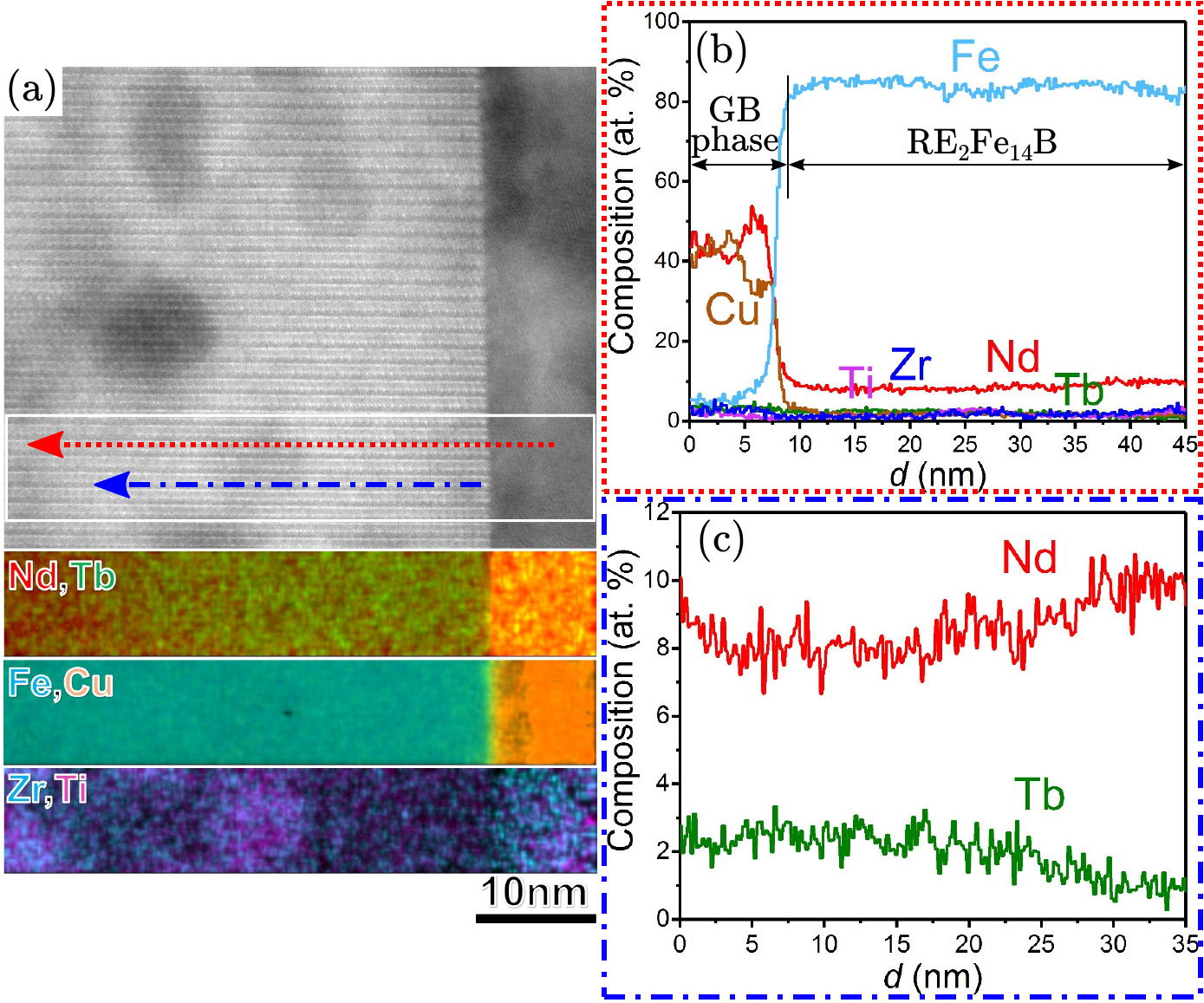}
	\caption{Composition of the GB phase of the LPBF and with Nd$_{50}$Tb$_{20}$Cu$_{30}$ diffusion-processed sample. (a) STEM-HAADF image and STEM-EDS elemental mapping images of Nd,Tb; Fe,Cu; Zr,Ti. (b) Line scan of the composition along the red doted line. (c) Line scan of the composition along the blue dotdashed line.}
	\label{fig:nd_tb_cu_diffusion_composition}
\end{figure}

Remaining questions are what is the chemical composition of the Nd-rich intergranular phase and what is the Tb content at the surface of Nd$_{2}$Fe$_{14}$B grains? Fig.~\ref{fig:nd_tb_cu_diffusion_composition}(a) shows high magnification HAADF-STEM image obtained from the interface of a grain boundary phase with Nd$_{2}$Fe$_{14}$B grain. STEM-EDS maps of Nd, Tb, Fe, Cu, Zr, and Ti is shown in this figure, obtained from marked box in HAADF-STEM image. Composition line profile obtained from red arrow is shown in Fig.~\ref{fig:nd_tb_cu_diffusion_composition}(b) indicating the grain boundary phase does not contain any ferromagnetic elements and is enriched with Nd and Cu. In order to clarify the Tb content at the surface of Nd$_{2}$Fe$_{14}$B grains, composition line profile of Nd and Tb obtained from blue arrow, marked in (a), is shown in Fig.~\ref{fig:nd_tb_cu_diffusion_composition}(c) showing the surface 2:14:1 grains contain $~2$~at.\% of Tb. Enrichment of Tb at the surface of $2$:$14$:$1$ grains increases magnetocrystalline anisotropy field at the interfaces hindering nucleation of reversed domains at the interfaces resulting an increase in the coercivity. Unlike conventional Dy or Tb vapor diffusion process in Nd-Fe-B sintered magnets that require elevated temperature diffusion process of above $850$~$^\circ$C, Nd-Tb-Cu diffusion process was carried out at low temperature of $650$~$^\circ$C resulting formation of Tb-rich shell without substantial grain growth of Nd$_{2}$Fe$_{14}$B phase. This was also reported previously in the hot-deformed Nd-Fe-B magnet \cite{li2018coercivity}

\section*{Conclusion}
Laser powder bed fusion (LPBF) is a well established additive manufacturing (AM) method for the production of metal high performance metallic parts. Recently, several research groups work on AM methods for hard or soft magnetic objects. However, low coercivity of Nd-Fe-B magnets processed with LPBF has been one of the draw backs of these magnets.

The commercial isotropic NdFeB powder MQP-S from Magnetquench meets the powder requirements  for the LPBF process. For an efficient diffusion of the low melting point alloy, the powder is only sintered and not fully melted during the LPBF process. The printing parameters are optimized in order to produce mecchanically stable parts with a similar volumetric mass density as the powder tab density. The coercivity of the sintered parts is with $0.65$~T slightly lower than the coercivity of the powder ($0.88$~T). This low coercivity is a result of the absence of the Nd-rich grain boundary phase which separates nano-sized Nd$_{2}$Fe$_{14}$B grains as well as of the existence of soft $\alpha$-Fe phase in the microstructure.

For the coercivity enhancement of the printed parts, low melting point melt-spun ribbons with compositions of Nd$_{70}$Cu$_{30}$, Nd$_{80}$Cu$_{20}$, Nd$_{60}$Al$_{10}$Ni$_{10}$Cu$_{20}$, and Nd$_{50}$Tb$_{20}$Cu$_{20}$ are investigated. Diffusion of Nd-Cu alloys and Nd-Al-Ni-Cu alloys increases the coercivity from $0.65$~T to $1.0$~T. However, use of Nd-Tb-Cu as the diffusion source results in achieving a larger coercivity of $1.5$~T in the samples. After diffusion process, the samples showed a full density of above $7.5$~g/cm$^3$. Enrichment of Tb at the surface of Nd$_{2}$Fe$_{14}$B grains increases magnetocrystalline anisotropy field at the interfaces hindering nucleation of reversed domains at the interfaces resulting an increase in the coercivity.

\section*{Acknowledgment}
The support from CD-Laboratory AMSEN (financed by the Austrian Federal Ministry of Economy, Family and Youth, the National Foundation for Research, Technology and Development) is acknowledged.


\begin{thebibliography}{28}%
\makeatletter
\providecommand \@ifxundefined [1]{%
 \@ifx{#1\undefined}
}%
\providecommand \@ifnum [1]{%
 \ifnum #1\expandafter \@firstoftwo
 \else \expandafter \@secondoftwo
 \fi
}%
\providecommand \@ifx [1]{%
 \ifx #1\expandafter \@firstoftwo
 \else \expandafter \@secondoftwo
 \fi
}%
\providecommand \natexlab [1]{#1}%
\providecommand \enquote  [1]{``#1''}%
\providecommand \bibnamefont  [1]{#1}%
\providecommand \bibfnamefont [1]{#1}%
\providecommand \citenamefont [1]{#1}%
\providecommand \href@noop [0]{\@secondoftwo}%
\providecommand \href [0]{\begingroup \@sanitize@url \@href}%
\providecommand \@href[1]{\@@startlink{#1}\@@href}%
\providecommand \@@href[1]{\endgroup#1\@@endlink}%
\providecommand \@sanitize@url [0]{\catcode `\\12\catcode `\$12\catcode
  `\&12\catcode `\#12\catcode `\^12\catcode `\_12\catcode `\%12\relax}%
\providecommand \@@startlink[1]{}%
\providecommand \@@endlink[0]{}%
\providecommand \url  [0]{\begingroup\@sanitize@url \@url }%
\providecommand \@url [1]{\endgroup\@href {#1}{\urlprefix }}%
\providecommand \urlprefix  [0]{URL }%
\providecommand \Eprint [0]{\href }%
\providecommand \doibase [0]{http://dx.doi.org/}%
\providecommand \selectlanguage [0]{\@gobble}%
\providecommand \bibinfo  [0]{\@secondoftwo}%
\providecommand \bibfield  [0]{\@secondoftwo}%
\providecommand \translation [1]{[#1]}%
\providecommand \BibitemOpen [0]{}%
\providecommand \bibitemStop [0]{}%
\providecommand \bibitemNoStop [0]{.\EOS\space}%
\providecommand \EOS [0]{\spacefactor3000\relax}%
\providecommand \BibitemShut  [1]{\csname bibitem#1\endcsname}%
\let\auto@bib@innerbib\@empty
\bibitem [{\citenamefont {Huber}\ \emph {et~al.}(2016)\citenamefont {Huber},
  \citenamefont {Abert}, \citenamefont {Bruckner}, \citenamefont {Groenefeld},
  \citenamefont {Muthsam}, \citenamefont {Schuschnigg}, \citenamefont {Sirak},
  \citenamefont {Thanhoffer}, \citenamefont {Teliban}, \citenamefont {Vogler},
  \citenamefont {Windl},\ and\ \citenamefont {Suess}}]{pub_16_1_apl}%
  \BibitemOpen
  \bibfield  {author} {\bibinfo {author} {\bibfnamefont {C.}~\bibnamefont
  {Huber}}, \bibinfo {author} {\bibfnamefont {C.}~\bibnamefont {Abert}},
  \bibinfo {author} {\bibfnamefont {F.}~\bibnamefont {Bruckner}}, \bibinfo
  {author} {\bibfnamefont {M.}~\bibnamefont {Groenefeld}}, \bibinfo {author}
  {\bibfnamefont {O.}~\bibnamefont {Muthsam}}, \bibinfo {author} {\bibfnamefont
  {S.}~\bibnamefont {Schuschnigg}}, \bibinfo {author} {\bibfnamefont
  {K.}~\bibnamefont {Sirak}}, \bibinfo {author} {\bibfnamefont
  {R.}~\bibnamefont {Thanhoffer}}, \bibinfo {author} {\bibfnamefont
  {I.}~\bibnamefont {Teliban}}, \bibinfo {author} {\bibfnamefont
  {C.}~\bibnamefont {Vogler}}, \bibinfo {author} {\bibfnamefont
  {R.}~\bibnamefont {Windl}}, \ and\ \bibinfo {author} {\bibfnamefont
  {D.}~\bibnamefont {Suess}},\ }\href@noop {} {\bibfield  {journal} {\bibinfo
  {journal} {Applied Physics Letters}\ }\textbf {\bibinfo {volume} {109}},\
  \bibinfo {pages} {162401} (\bibinfo {year} {2016})}\BibitemShut {NoStop}%
\bibitem [{\citenamefont {Huber}\ \emph
  {et~al.}(2017{\natexlab{a}})\citenamefont {Huber}, \citenamefont {Abert},
  \citenamefont {Bruckner}, \citenamefont {Pfaff}, \citenamefont {Kriwet},
  \citenamefont {Groenefeld}, \citenamefont {Teliban}, \citenamefont {Vogler},\
  and\ \citenamefont {Suess}}]{pub_17_2}%
  \BibitemOpen
  \bibfield  {author} {\bibinfo {author} {\bibfnamefont {C.}~\bibnamefont
  {Huber}}, \bibinfo {author} {\bibfnamefont {C.}~\bibnamefont {Abert}},
  \bibinfo {author} {\bibfnamefont {F.}~\bibnamefont {Bruckner}}, \bibinfo
  {author} {\bibfnamefont {C.}~\bibnamefont {Pfaff}}, \bibinfo {author}
  {\bibfnamefont {J.}~\bibnamefont {Kriwet}}, \bibinfo {author} {\bibfnamefont
  {M.}~\bibnamefont {Groenefeld}}, \bibinfo {author} {\bibfnamefont
  {I.}~\bibnamefont {Teliban}}, \bibinfo {author} {\bibfnamefont
  {C.}~\bibnamefont {Vogler}}, \ and\ \bibinfo {author} {\bibfnamefont
  {D.}~\bibnamefont {Suess}},\ }\href {\doibase 10.1063/1.4997441} {\bibfield
  {journal} {\bibinfo  {journal} {Journal of Applied Physics}\ }\textbf
  {\bibinfo {volume} {122}},\ \bibinfo {pages} {053904} (\bibinfo {year}
  {2017}{\natexlab{a}})}\BibitemShut {NoStop}%
\bibitem [{\citenamefont {Huber}\ \emph
  {et~al.}(2017{\natexlab{b}})\citenamefont {Huber}, \citenamefont {Abert},
  \citenamefont {Bruckner}, \citenamefont {Groenefeld}, \citenamefont
  {Schuschnigg}, \citenamefont {Teliban}, \citenamefont {Vogler}, \citenamefont
  {Wautischer}, \citenamefont {Windl},\ and\ \citenamefont {Suess}}]{pub_17_1}%
  \BibitemOpen
  \bibfield  {author} {\bibinfo {author} {\bibfnamefont {C.}~\bibnamefont
  {Huber}}, \bibinfo {author} {\bibfnamefont {C.}~\bibnamefont {Abert}},
  \bibinfo {author} {\bibfnamefont {F.}~\bibnamefont {Bruckner}}, \bibinfo
  {author} {\bibfnamefont {M.}~\bibnamefont {Groenefeld}}, \bibinfo {author}
  {\bibfnamefont {S.}~\bibnamefont {Schuschnigg}}, \bibinfo {author}
  {\bibfnamefont {I.}~\bibnamefont {Teliban}}, \bibinfo {author} {\bibfnamefont
  {C.}~\bibnamefont {Vogler}}, \bibinfo {author} {\bibfnamefont
  {G.}~\bibnamefont {Wautischer}}, \bibinfo {author} {\bibfnamefont
  {R.}~\bibnamefont {Windl}}, \ and\ \bibinfo {author} {\bibfnamefont
  {D.}~\bibnamefont {Suess}},\ }\href@noop {} {\bibfield  {journal} {\bibinfo
  {journal} {Scientific Reports}\ }\textbf {\bibinfo {volume} {7}},\ \bibinfo
  {pages} {9419} (\bibinfo {year} {2017}{\natexlab{b}})}\BibitemShut {NoStop}%
\bibitem [{\citenamefont {Ortner}\ \emph {et~al.}(2017)\citenamefont {Ortner},
  \citenamefont {Huber}, \citenamefont {Vollert}, \citenamefont {Pilz},\ and\
  \citenamefont {S{\"u}ss}}]{ortner2017application}%
  \BibitemOpen
  \bibfield  {author} {\bibinfo {author} {\bibfnamefont {M.}~\bibnamefont
  {Ortner}}, \bibinfo {author} {\bibfnamefont {C.}~\bibnamefont {Huber}},
  \bibinfo {author} {\bibfnamefont {N.}~\bibnamefont {Vollert}}, \bibinfo
  {author} {\bibfnamefont {J.}~\bibnamefont {Pilz}}, \ and\ \bibinfo {author}
  {\bibfnamefont {D.}~\bibnamefont {S{\"u}ss}},\ }in\ \href@noop {} {\emph
  {\bibinfo {booktitle} {SENSORS, 2017 IEEE}}}\ (\bibinfo {organization}
  {IEEE},\ \bibinfo {year} {2017})\ pp.\ \bibinfo {pages} {1--3}\BibitemShut
  {NoStop}%
\bibitem [{\citenamefont {Li}\ \emph {et~al.}(2016)\citenamefont {Li},
  \citenamefont {Tirado}, \citenamefont {Nlebedim}, \citenamefont {Rios},
  \citenamefont {Post}, \citenamefont {Kunc}, \citenamefont {Lowden},
  \citenamefont {Lara-Curzio}, \citenamefont {Fredette}, \citenamefont
  {Ormerod}, \citenamefont {Lograsso},\ and\ \citenamefont
  {Paranthaman}}]{baam}%
  \BibitemOpen
  \bibfield  {author} {\bibinfo {author} {\bibfnamefont {L.}~\bibnamefont
  {Li}}, \bibinfo {author} {\bibfnamefont {A.}~\bibnamefont {Tirado}}, \bibinfo
  {author} {\bibfnamefont {I.~C.}\ \bibnamefont {Nlebedim}}, \bibinfo {author}
  {\bibfnamefont {O.}~\bibnamefont {Rios}}, \bibinfo {author} {\bibfnamefont
  {B.}~\bibnamefont {Post}}, \bibinfo {author} {\bibfnamefont {V.}~\bibnamefont
  {Kunc}}, \bibinfo {author} {\bibfnamefont {R.~R.}\ \bibnamefont {Lowden}},
  \bibinfo {author} {\bibfnamefont {E.}~\bibnamefont {Lara-Curzio}}, \bibinfo
  {author} {\bibfnamefont {R.}~\bibnamefont {Fredette}}, \bibinfo {author}
  {\bibfnamefont {J.}~\bibnamefont {Ormerod}}, \bibinfo {author} {\bibfnamefont
  {T.~A.}\ \bibnamefont {Lograsso}}, \ and\ \bibinfo {author} {\bibfnamefont
  {M.~P.}\ \bibnamefont {Paranthaman}},\ }\href@noop {} {\bibfield  {journal}
  {\bibinfo  {journal} {Scientific Reports}\ }\textbf {\bibinfo {volume} {6}},\
  \bibinfo {pages} {36212} (\bibinfo {year} {2016})}\BibitemShut {NoStop}%
\bibitem [{\citenamefont {von Petersdorff-Campen}\ \emph
  {et~al.}(2018)\citenamefont {von Petersdorff-Campen}, \citenamefont
  {Hauswirth}, \citenamefont {Carpenter}, \citenamefont {Hagmann},
  \citenamefont {Bo{\"e}s}, \citenamefont {Schmid~Daners}, \citenamefont
  {Penner},\ and\ \citenamefont {Meboldt}}]{von20183d}%
  \BibitemOpen
  \bibfield  {author} {\bibinfo {author} {\bibfnamefont {K.}~\bibnamefont {von
  Petersdorff-Campen}}, \bibinfo {author} {\bibfnamefont {Y.}~\bibnamefont
  {Hauswirth}}, \bibinfo {author} {\bibfnamefont {J.}~\bibnamefont
  {Carpenter}}, \bibinfo {author} {\bibfnamefont {A.}~\bibnamefont {Hagmann}},
  \bibinfo {author} {\bibfnamefont {S.}~\bibnamefont {Bo{\"e}s}}, \bibinfo
  {author} {\bibfnamefont {M.}~\bibnamefont {Schmid~Daners}}, \bibinfo {author}
  {\bibfnamefont {D.}~\bibnamefont {Penner}}, \ and\ \bibinfo {author}
  {\bibfnamefont {M.}~\bibnamefont {Meboldt}},\ }\href@noop {} {\bibfield
  {journal} {\bibinfo  {journal} {Applied Sciences}\ }\textbf {\bibinfo
  {volume} {8}},\ \bibinfo {pages} {1275} (\bibinfo {year} {2018})}\BibitemShut
  {NoStop}%
\bibitem [{\citenamefont {Paranthaman}\ \emph {et~al.}(2016)\citenamefont
  {Paranthaman}, \citenamefont {Shafer}, \citenamefont {Elliott}, \citenamefont
  {Siddel}, \citenamefont {McGuire}, \citenamefont {Springfield}, \citenamefont
  {Martin}, \citenamefont {Fredette},\ and\ \citenamefont
  {Ormerod}}]{binder_jetting}%
  \BibitemOpen
  \bibfield  {author} {\bibinfo {author} {\bibfnamefont {M.~P.}\ \bibnamefont
  {Paranthaman}}, \bibinfo {author} {\bibfnamefont {C.~S.}\ \bibnamefont
  {Shafer}}, \bibinfo {author} {\bibfnamefont {A.~M.}\ \bibnamefont {Elliott}},
  \bibinfo {author} {\bibfnamefont {D.~H.}\ \bibnamefont {Siddel}}, \bibinfo
  {author} {\bibfnamefont {M.~A.}\ \bibnamefont {McGuire}}, \bibinfo {author}
  {\bibfnamefont {R.~M.}\ \bibnamefont {Springfield}}, \bibinfo {author}
  {\bibfnamefont {J.}~\bibnamefont {Martin}}, \bibinfo {author} {\bibfnamefont
  {R.}~\bibnamefont {Fredette}}, \ and\ \bibinfo {author} {\bibfnamefont
  {J.}~\bibnamefont {Ormerod}},\ }\href@noop {} {\bibfield  {journal} {\bibinfo
   {journal} {JOM}\ }\textbf {\bibinfo {volume} {68}},\ \bibinfo {pages} {1978}
  (\bibinfo {year} {2016})}\BibitemShut {NoStop}%
\bibitem [{\citenamefont {Li}\ \emph {et~al.}(2017)\citenamefont {Li},
  \citenamefont {Tirado}, \citenamefont {Conner}, \citenamefont {Chi},
  \citenamefont {Elliott}, \citenamefont {Rios}, \citenamefont {Zhou},\ and\
  \citenamefont {Paranthaman}}]{li2017novel}%
  \BibitemOpen
  \bibfield  {author} {\bibinfo {author} {\bibfnamefont {L.}~\bibnamefont
  {Li}}, \bibinfo {author} {\bibfnamefont {A.}~\bibnamefont {Tirado}}, \bibinfo
  {author} {\bibfnamefont {B.~S.}\ \bibnamefont {Conner}}, \bibinfo {author}
  {\bibfnamefont {M.}~\bibnamefont {Chi}}, \bibinfo {author} {\bibfnamefont
  {A.~M.}\ \bibnamefont {Elliott}}, \bibinfo {author} {\bibfnamefont
  {O.}~\bibnamefont {Rios}}, \bibinfo {author} {\bibfnamefont {H.}~\bibnamefont
  {Zhou}}, \ and\ \bibinfo {author} {\bibfnamefont {M.~P.}\ \bibnamefont
  {Paranthaman}},\ }\href@noop {} {\bibfield  {journal} {\bibinfo  {journal}
  {Journal of Magnetism and Magnetic Materials}\ }\textbf {\bibinfo {volume}
  {438}},\ \bibinfo {pages} {163} (\bibinfo {year} {2017})}\BibitemShut
  {NoStop}%
\bibitem [{\citenamefont {Kruth}\ \emph {et~al.}(2004)\citenamefont {Kruth},
  \citenamefont {Froyen}, \citenamefont {Vaerenbergh}, \citenamefont
  {Mercelis}, \citenamefont {Rombouts},\ and\ \citenamefont {Lauwers}}]{slm}%
  \BibitemOpen
  \bibfield  {author} {\bibinfo {author} {\bibfnamefont {J.}~\bibnamefont
  {Kruth}}, \bibinfo {author} {\bibfnamefont {L.}~\bibnamefont {Froyen}},
  \bibinfo {author} {\bibfnamefont {J.~V.}\ \bibnamefont {Vaerenbergh}},
  \bibinfo {author} {\bibfnamefont {P.}~\bibnamefont {Mercelis}}, \bibinfo
  {author} {\bibfnamefont {M.}~\bibnamefont {Rombouts}}, \ and\ \bibinfo
  {author} {\bibfnamefont {B.}~\bibnamefont {Lauwers}},\ }\href@noop {}
  {\bibfield  {journal} {\bibinfo  {journal} {Journal of Materials Processing
  Technology}\ }\textbf {\bibinfo {volume} {149}},\ \bibinfo {pages} {616 }
  (\bibinfo {year} {2004})},\ \bibinfo {note} {14th Interntaional Symposium on
  Electromachining (ISEM XIV)}\BibitemShut {NoStop}%
\bibitem [{\citenamefont {Attar}\ \emph {et~al.}(2015)\citenamefont {Attar},
  \citenamefont {Prashanth}, \citenamefont {Zhang}, \citenamefont {Calin},
  \citenamefont {Okulov}, \citenamefont {Scudino}, \citenamefont {Yang},\ and\
  \citenamefont {Eckert}}]{attar2015effect}%
  \BibitemOpen
  \bibfield  {author} {\bibinfo {author} {\bibfnamefont {H.}~\bibnamefont
  {Attar}}, \bibinfo {author} {\bibfnamefont {K.~G.}\ \bibnamefont
  {Prashanth}}, \bibinfo {author} {\bibfnamefont {L.-C.}\ \bibnamefont
  {Zhang}}, \bibinfo {author} {\bibfnamefont {M.}~\bibnamefont {Calin}},
  \bibinfo {author} {\bibfnamefont {I.~V.}\ \bibnamefont {Okulov}}, \bibinfo
  {author} {\bibfnamefont {S.}~\bibnamefont {Scudino}}, \bibinfo {author}
  {\bibfnamefont {C.}~\bibnamefont {Yang}}, \ and\ \bibinfo {author}
  {\bibfnamefont {J.}~\bibnamefont {Eckert}},\ }\href@noop {} {\bibfield
  {journal} {\bibinfo  {journal} {Journal of Materials Science \& Technology}\
  }\textbf {\bibinfo {volume} {31}},\ \bibinfo {pages} {1001} (\bibinfo {year}
  {2015})}\BibitemShut {NoStop}%
\bibitem [{\citenamefont {Huber}\ \emph {et~al.}(2018)\citenamefont {Huber},
  \citenamefont {Goertler}, \citenamefont {Abert}, \citenamefont {Bruckner},
  \citenamefont {Groenefeld}, \citenamefont {Teliban},\ and\ \citenamefont
  {Suess}}]{huber2018additive}%
  \BibitemOpen
  \bibfield  {author} {\bibinfo {author} {\bibfnamefont {C.}~\bibnamefont
  {Huber}}, \bibinfo {author} {\bibfnamefont {M.}~\bibnamefont {Goertler}},
  \bibinfo {author} {\bibfnamefont {C.}~\bibnamefont {Abert}}, \bibinfo
  {author} {\bibfnamefont {F.}~\bibnamefont {Bruckner}}, \bibinfo {author}
  {\bibfnamefont {M.}~\bibnamefont {Groenefeld}}, \bibinfo {author}
  {\bibfnamefont {I.}~\bibnamefont {Teliban}}, \ and\ \bibinfo {author}
  {\bibfnamefont {D.}~\bibnamefont {Suess}},\ }\href@noop {} {\bibfield
  {journal} {\bibinfo  {journal} {Scientific reports}\ }\textbf {\bibinfo
  {volume} {8}},\ \bibinfo {pages} {14651} (\bibinfo {year}
  {2018})}\BibitemShut {NoStop}%
\bibitem [{\citenamefont {Mikler}\ \emph {et~al.}(2017)\citenamefont {Mikler},
  \citenamefont {Chaudhary}, \citenamefont {Borkar}, \citenamefont {Soni},
  \citenamefont {Jaeger}, \citenamefont {Chen}, \citenamefont {Contieri},
  \citenamefont {Ramanujan},\ and\ \citenamefont {Banerjee}}]{mikler2017laser}%
  \BibitemOpen
  \bibfield  {author} {\bibinfo {author} {\bibfnamefont {C.}~\bibnamefont
  {Mikler}}, \bibinfo {author} {\bibfnamefont {V.}~\bibnamefont {Chaudhary}},
  \bibinfo {author} {\bibfnamefont {T.}~\bibnamefont {Borkar}}, \bibinfo
  {author} {\bibfnamefont {V.}~\bibnamefont {Soni}}, \bibinfo {author}
  {\bibfnamefont {D.}~\bibnamefont {Jaeger}}, \bibinfo {author} {\bibfnamefont
  {X.}~\bibnamefont {Chen}}, \bibinfo {author} {\bibfnamefont {R.}~\bibnamefont
  {Contieri}}, \bibinfo {author} {\bibfnamefont {R.}~\bibnamefont {Ramanujan}},
  \ and\ \bibinfo {author} {\bibfnamefont {R.}~\bibnamefont {Banerjee}},\
  }\href@noop {} {\bibfield  {journal} {\bibinfo  {journal} {JOM}\ }\textbf
  {\bibinfo {volume} {69}},\ \bibinfo {pages} {532} (\bibinfo {year}
  {2017})}\BibitemShut {NoStop}%
\bibitem [{\citenamefont {Zhang}\ \emph {et~al.}(2013)\citenamefont {Zhang},
  \citenamefont {Fenineche}, \citenamefont {Liao},\ and\ \citenamefont
  {Coddet}}]{zhang2013magnetic}%
  \BibitemOpen
  \bibfield  {author} {\bibinfo {author} {\bibfnamefont {B.}~\bibnamefont
  {Zhang}}, \bibinfo {author} {\bibfnamefont {N.-E.}\ \bibnamefont
  {Fenineche}}, \bibinfo {author} {\bibfnamefont {H.}~\bibnamefont {Liao}}, \
  and\ \bibinfo {author} {\bibfnamefont {C.}~\bibnamefont {Coddet}},\
  }\href@noop {} {\bibfield  {journal} {\bibinfo  {journal} {Journal of
  Magnetism and Magnetic Materials}\ }\textbf {\bibinfo {volume} {336}},\
  \bibinfo {pages} {49} (\bibinfo {year} {2013})}\BibitemShut {NoStop}%
\bibitem [{\citenamefont {White}\ \emph {et~al.}(2017)\citenamefont {White},
  \citenamefont {Kassen}, \citenamefont {Simsek}, \citenamefont {Tang},
  \citenamefont {Ott},\ and\ \citenamefont {Anderson}}]{white2017net}%
  \BibitemOpen
  \bibfield  {author} {\bibinfo {author} {\bibfnamefont {E.~M.~H.}\
  \bibnamefont {White}}, \bibinfo {author} {\bibfnamefont {A.~G.}\ \bibnamefont
  {Kassen}}, \bibinfo {author} {\bibfnamefont {E.}~\bibnamefont {Simsek}},
  \bibinfo {author} {\bibfnamefont {W.}~\bibnamefont {Tang}}, \bibinfo {author}
  {\bibfnamefont {R.~T.}\ \bibnamefont {Ott}}, \ and\ \bibinfo {author}
  {\bibfnamefont {I.~E.}\ \bibnamefont {Anderson}},\ }\href@noop {} {\bibfield
  {journal} {\bibinfo  {journal} {IEEE Transactions on Magnetics}\ }\textbf
  {\bibinfo {volume} {53}},\ \bibinfo {pages} {1} (\bibinfo {year}
  {2017})}\BibitemShut {NoStop}%
\bibitem [{\citenamefont {Ja{\'c}imovi{\'c}}\ \emph {et~al.}(2017)\citenamefont
  {Ja{\'c}imovi{\'c}}, \citenamefont {Binda}, \citenamefont {Herrmann},
  \citenamefont {Greuter}, \citenamefont {Genta}, \citenamefont {Calvo},
  \citenamefont {Tom{\v{s}}e},\ and\ \citenamefont {Simon}}]{lasersinter_mag}%
  \BibitemOpen
  \bibfield  {author} {\bibinfo {author} {\bibfnamefont {J.}~\bibnamefont
  {Ja{\'c}imovi{\'c}}}, \bibinfo {author} {\bibfnamefont {F.}~\bibnamefont
  {Binda}}, \bibinfo {author} {\bibfnamefont {L.~G.}\ \bibnamefont {Herrmann}},
  \bibinfo {author} {\bibfnamefont {F.}~\bibnamefont {Greuter}}, \bibinfo
  {author} {\bibfnamefont {J.}~\bibnamefont {Genta}}, \bibinfo {author}
  {\bibfnamefont {M.}~\bibnamefont {Calvo}}, \bibinfo {author} {\bibfnamefont
  {T.}~\bibnamefont {Tom{\v{s}}e}}, \ and\ \bibinfo {author} {\bibfnamefont
  {R.~A.}\ \bibnamefont {Simon}},\ }\href@noop {} {\bibfield  {journal}
  {\bibinfo  {journal} {Advanced Engineering Materials}\ } (\bibinfo {year}
  {2017})}\BibitemShut {NoStop}%
\bibitem [{\citenamefont {Urban}, \citenamefont {Huber},\ and\ \citenamefont
  {Franke}(2017)}]{urban2017influences}%
  \BibitemOpen
  \bibfield  {author} {\bibinfo {author} {\bibfnamefont {N.}~\bibnamefont
  {Urban}}, \bibinfo {author} {\bibfnamefont {F.}~\bibnamefont {Huber}}, \ and\
  \bibinfo {author} {\bibfnamefont {J.}~\bibnamefont {Franke}},\ }in\
  \href@noop {} {\emph {\bibinfo {booktitle} {Electric Drives Production
  Conference (EDPC), 2017 7th International}}}\ (\bibinfo {organization}
  {IEEE},\ \bibinfo {year} {2017})\ pp.\ \bibinfo {pages} {1--5}\BibitemShut
  {NoStop}%
\bibitem [{\citenamefont {Kanekiyo}, \citenamefont {Miyoshi},\ and\
  \citenamefont {Hirosawa}(2004)}]{kanekiyo2004nanocomposite}%
  \BibitemOpen
  \bibfield  {author} {\bibinfo {author} {\bibfnamefont {H.}~\bibnamefont
  {Kanekiyo}}, \bibinfo {author} {\bibfnamefont {T.}~\bibnamefont {Miyoshi}}, \
  and\ \bibinfo {author} {\bibfnamefont {S.}~\bibnamefont {Hirosawa}},\
  }\href@noop {} {\enquote {\bibinfo {title} {Nanocomposite magnet and method
  for producing same},}\ } (\bibinfo {year} {2004}),\ \bibinfo {note} {uS
  Patent 6,790,296}\BibitemShut {NoStop}%
\bibitem [{\citenamefont {Fischer}\ \emph {et~al.}(1996)\citenamefont
  {Fischer}, \citenamefont {Schrefl}, \citenamefont {Kronm{\"u}ller},\ and\
  \citenamefont {Fidler}}]{fischer1996grain}%
  \BibitemOpen
  \bibfield  {author} {\bibinfo {author} {\bibfnamefont {R.}~\bibnamefont
  {Fischer}}, \bibinfo {author} {\bibfnamefont {T.}~\bibnamefont {Schrefl}},
  \bibinfo {author} {\bibfnamefont {H.}~\bibnamefont {Kronm{\"u}ller}}, \ and\
  \bibinfo {author} {\bibfnamefont {J.}~\bibnamefont {Fidler}},\ }\href@noop {}
  {\bibfield  {journal} {\bibinfo  {journal} {Journal of magnetism and magnetic
  materials}\ }\textbf {\bibinfo {volume} {153}},\ \bibinfo {pages} {35}
  (\bibinfo {year} {1996})}\BibitemShut {NoStop}%
\bibitem [{\citenamefont {Engelmann}, \citenamefont {Kim},\ and\ \citenamefont
  {Thomas}(1997)}]{engelmann1997microstructure}%
  \BibitemOpen
  \bibfield  {author} {\bibinfo {author} {\bibfnamefont {H.}~\bibnamefont
  {Engelmann}}, \bibinfo {author} {\bibfnamefont {A.}~\bibnamefont {Kim}}, \
  and\ \bibinfo {author} {\bibfnamefont {G.}~\bibnamefont {Thomas}},\
  }\href@noop {} {\bibfield  {journal} {\bibinfo  {journal} {Scripta
  materialia}\ }\textbf {\bibinfo {volume} {36}},\ \bibinfo {pages} {55}
  (\bibinfo {year} {1997})}\BibitemShut {NoStop}%
\bibitem [{\citenamefont {Kronm{\"u}ller}, \citenamefont {Durst},\ and\
  \citenamefont {Sagawa}(1988)}]{kronmuller1988analysis}%
  \BibitemOpen
  \bibfield  {author} {\bibinfo {author} {\bibfnamefont {H.}~\bibnamefont
  {Kronm{\"u}ller}}, \bibinfo {author} {\bibfnamefont {K.-D.}\ \bibnamefont
  {Durst}}, \ and\ \bibinfo {author} {\bibfnamefont {M.}~\bibnamefont
  {Sagawa}},\ }\href@noop {} {\bibfield  {journal} {\bibinfo  {journal}
  {Journal of magnetism and magnetic materials}\ }\textbf {\bibinfo {volume}
  {74}},\ \bibinfo {pages} {291} (\bibinfo {year} {1988})}\BibitemShut
  {NoStop}%
\bibitem [{\citenamefont {Hono}\ and\ \citenamefont
  {Sepehri-Amin}(2012)}]{hono2012strategy}%
  \BibitemOpen
  \bibfield  {author} {\bibinfo {author} {\bibfnamefont {K.}~\bibnamefont
  {Hono}}\ and\ \bibinfo {author} {\bibfnamefont {H.}~\bibnamefont
  {Sepehri-Amin}},\ }\href@noop {} {\bibfield  {journal} {\bibinfo  {journal}
  {Scripta Materialia}\ }\textbf {\bibinfo {volume} {67}},\ \bibinfo {pages}
  {530} (\bibinfo {year} {2012})}\BibitemShut {NoStop}%
\bibitem [{\citenamefont {Boltich}\ \emph {et~al.}(1985)\citenamefont
  {Boltich}, \citenamefont {Oswald}, \citenamefont {Huang}, \citenamefont
  {Hirosawa}, \citenamefont {Wallace},\ and\ \citenamefont
  {Burzo}}]{boltich1985magnetic}%
  \BibitemOpen
  \bibfield  {author} {\bibinfo {author} {\bibfnamefont {E.}~\bibnamefont
  {Boltich}}, \bibinfo {author} {\bibfnamefont {E.}~\bibnamefont {Oswald}},
  \bibinfo {author} {\bibfnamefont {M.}~\bibnamefont {Huang}}, \bibinfo
  {author} {\bibfnamefont {S.}~\bibnamefont {Hirosawa}}, \bibinfo {author}
  {\bibfnamefont {W.}~\bibnamefont {Wallace}}, \ and\ \bibinfo {author}
  {\bibfnamefont {E.}~\bibnamefont {Burzo}},\ }\href@noop {} {\bibfield
  {journal} {\bibinfo  {journal} {Journal of applied physics}\ }\textbf
  {\bibinfo {volume} {57}},\ \bibinfo {pages} {4106} (\bibinfo {year}
  {1985})}\BibitemShut {NoStop}%
\bibitem [{\citenamefont {Hirosawa}\ \emph {et~al.}(1986)\citenamefont
  {Hirosawa}, \citenamefont {Matsuura}, \citenamefont {Yamamoto}, \citenamefont
  {Fujimura}, \citenamefont {Sagawa},\ and\ \citenamefont
  {Yamauchi}}]{hirosawa1986magnetization}%
  \BibitemOpen
  \bibfield  {author} {\bibinfo {author} {\bibfnamefont {S.}~\bibnamefont
  {Hirosawa}}, \bibinfo {author} {\bibfnamefont {Y.}~\bibnamefont {Matsuura}},
  \bibinfo {author} {\bibfnamefont {H.}~\bibnamefont {Yamamoto}}, \bibinfo
  {author} {\bibfnamefont {S.}~\bibnamefont {Fujimura}}, \bibinfo {author}
  {\bibfnamefont {M.}~\bibnamefont {Sagawa}}, \ and\ \bibinfo {author}
  {\bibfnamefont {H.}~\bibnamefont {Yamauchi}},\ }\href@noop {} {\bibfield
  {journal} {\bibinfo  {journal} {Journal of applied physics}\ }\textbf
  {\bibinfo {volume} {59}},\ \bibinfo {pages} {873} (\bibinfo {year}
  {1986})}\BibitemShut {NoStop}%
\bibitem [{\citenamefont {Liu}\ \emph {et~al.}(2016)\citenamefont {Liu},
  \citenamefont {Akiya}, \citenamefont {Sepehri-Amin}, \citenamefont {Ohkubo},
  \citenamefont {Sakuma}, \citenamefont {Yano}, \citenamefont {Kato},\ and\
  \citenamefont {Hono}}]{liu2016coercivity}%
  \BibitemOpen
  \bibfield  {author} {\bibinfo {author} {\bibfnamefont {L.}~\bibnamefont
  {Liu}}, \bibinfo {author} {\bibfnamefont {T.}~\bibnamefont {Akiya}}, \bibinfo
  {author} {\bibfnamefont {H.}~\bibnamefont {Sepehri-Amin}}, \bibinfo {author}
  {\bibfnamefont {T.}~\bibnamefont {Ohkubo}}, \bibinfo {author} {\bibfnamefont
  {N.}~\bibnamefont {Sakuma}}, \bibinfo {author} {\bibfnamefont
  {M.}~\bibnamefont {Yano}}, \bibinfo {author} {\bibfnamefont {A.}~\bibnamefont
  {Kato}}, \ and\ \bibinfo {author} {\bibfnamefont {K.}~\bibnamefont {Hono}},\
  }\href@noop {} {\bibfield  {journal} {\bibinfo  {journal} {Journal of
  Magnetism and Magnetic Materials}\ }\textbf {\bibinfo {volume} {412}},\
  \bibinfo {pages} {234} (\bibinfo {year} {2016})}\BibitemShut {NoStop}%
\bibitem [{\citenamefont {Li}\ \emph {et~al.}(2018)\citenamefont {Li},
  \citenamefont {Liu}, \citenamefont {Sepehri-Amin}, \citenamefont {Tang},
  \citenamefont {Ohkubo}, \citenamefont {Sakuma}, \citenamefont {Shoji},
  \citenamefont {Kato}, \citenamefont {Schrefl},\ and\ \citenamefont
  {Hono}}]{li2018coercivity}%
  \BibitemOpen
  \bibfield  {author} {\bibinfo {author} {\bibfnamefont {J.}~\bibnamefont
  {Li}}, \bibinfo {author} {\bibfnamefont {L.}~\bibnamefont {Liu}}, \bibinfo
  {author} {\bibfnamefont {H.}~\bibnamefont {Sepehri-Amin}}, \bibinfo {author}
  {\bibfnamefont {X.}~\bibnamefont {Tang}}, \bibinfo {author} {\bibfnamefont
  {T.}~\bibnamefont {Ohkubo}}, \bibinfo {author} {\bibfnamefont
  {N.}~\bibnamefont {Sakuma}}, \bibinfo {author} {\bibfnamefont
  {T.}~\bibnamefont {Shoji}}, \bibinfo {author} {\bibfnamefont
  {A.}~\bibnamefont {Kato}}, \bibinfo {author} {\bibfnamefont {T.}~\bibnamefont
  {Schrefl}}, \ and\ \bibinfo {author} {\bibfnamefont {K.}~\bibnamefont
  {Hono}},\ }\href@noop {} {\bibfield  {journal} {\bibinfo  {journal} {Acta
  Materialia}\ }\textbf {\bibinfo {volume} {161}},\ \bibinfo {pages} {171}
  (\bibinfo {year} {2018})}\BibitemShut {NoStop}%
\bibitem [{\citenamefont {https://mqitechnology.com/}()}]{mqp}%
  \BibitemOpen
  \bibfield  {author} {\bibinfo {author} {\bibnamefont
  {https://mqitechnology.com/}},\ }\href
  {https://mqitechnology.com/product/mqp-s-11-9-20001/?pname=MQP-S-11-9&ppart=20001}
  {}\BibitemShut {NoStop}%
\bibitem [{\citenamefont {DebRoy}\ \emph {et~al.}(2018)\citenamefont {DebRoy},
  \citenamefont {Wei}, \citenamefont {Zuback}, \citenamefont {Mukherjee},
  \citenamefont {Elmer}, \citenamefont {Milewski}, \citenamefont {Beese},
  \citenamefont {Wilson-Heid}, \citenamefont {De},\ and\ \citenamefont
  {Zhang}}]{debroy2018additive}%
  \BibitemOpen
  \bibfield  {author} {\bibinfo {author} {\bibfnamefont {T.}~\bibnamefont
  {DebRoy}}, \bibinfo {author} {\bibfnamefont {H.}~\bibnamefont {Wei}},
  \bibinfo {author} {\bibfnamefont {J.}~\bibnamefont {Zuback}}, \bibinfo
  {author} {\bibfnamefont {T.}~\bibnamefont {Mukherjee}}, \bibinfo {author}
  {\bibfnamefont {J.}~\bibnamefont {Elmer}}, \bibinfo {author} {\bibfnamefont
  {J.}~\bibnamefont {Milewski}}, \bibinfo {author} {\bibfnamefont
  {A.}~\bibnamefont {Beese}}, \bibinfo {author} {\bibfnamefont
  {A.}~\bibnamefont {Wilson-Heid}}, \bibinfo {author} {\bibfnamefont
  {A.}~\bibnamefont {De}}, \ and\ \bibinfo {author} {\bibfnamefont
  {W.}~\bibnamefont {Zhang}},\ }\href@noop {} {\bibfield  {journal} {\bibinfo
  {journal} {Progress in Materials Science}\ }\textbf {\bibinfo {volume}
  {92}},\ \bibinfo {pages} {112} (\bibinfo {year} {2018})}\BibitemShut
  {NoStop}%
\bibitem [{\citenamefont {Kolb}\ \emph {et~al.}(2016)\citenamefont {Kolb},
  \citenamefont {Huber}, \citenamefont {Akbulut}, \citenamefont {Donocik},
  \citenamefont {Urban}, \citenamefont {Maurer},\ and\ \citenamefont
  {Franke}}]{kolb2016laser}%
  \BibitemOpen
  \bibfield  {author} {\bibinfo {author} {\bibfnamefont {T.}~\bibnamefont
  {Kolb}}, \bibinfo {author} {\bibfnamefont {F.}~\bibnamefont {Huber}},
  \bibinfo {author} {\bibfnamefont {B.}~\bibnamefont {Akbulut}}, \bibinfo
  {author} {\bibfnamefont {C.}~\bibnamefont {Donocik}}, \bibinfo {author}
  {\bibfnamefont {N.}~\bibnamefont {Urban}}, \bibinfo {author} {\bibfnamefont
  {D.}~\bibnamefont {Maurer}}, \ and\ \bibinfo {author} {\bibfnamefont
  {J.}~\bibnamefont {Franke}},\ }in\ \href@noop {} {\emph {\bibinfo {booktitle}
  {Electric Drives Production Conference (EDPC), 2016 6th International}}}\
  (\bibinfo {organization} {IEEE},\ \bibinfo {year} {2016})\ pp.\ \bibinfo
  {pages} {34--40}\BibitemShut {NoStop}%
\end{thebibliography}

%

\end{document}